\documentstyle{mn}
\def\refitem{\par\parskip 0pt\noindent\hangindent 20pt}
\def \lk {L_{\rm kin}}
\def \lr {L_{\rm rot}}
\def \ll {L_{\rm BLR}}

\def \li {L_{\rm ion}}

\def \lha {L_{\rm H\alpha}}
\def \lla {L_{\rm Ly \alpha}}
\def \lha {L_{\rm H\alpha}}

\title[Jets and accretion processes in AGN] 
{Jets and accretion processes in Active Galactic Nuclei: further clues}

\author[A. Celotti, P. Padovani, G. Ghisellini]
{A. Celotti$^1$, P. Padovani$^2$, G. Ghisellini$^3$\\ 
$^1$ S.I.S.S.A., via Beirut 2/4, 34014 Trieste, Italy \\ 
$^2$ Dipartimento di Fisica, II Universit\`a di Roma ``Tor Vergata'', 
                  Via della Ricerca Scientifica 1, I-00133 Roma, Italy\\ 
$^3$ Osservatorio Astronomico di Brera--Merate, Via Bianchi 46, I-22055
                                                 Merate (Lecco), Italy}

\date{Received ***; in original form ***} 
       
\begin{document} 
\maketitle 
\begin{abstract} 

\noindent We present evidence in favour of a link between the luminosity
radiatively dissipated in the central engine of radio--loud Active
Galactic Nuclei and the kinetic power in their jets.  This piece of
evidence is based on the relation we find between the luminosity in broad
emission lines and the kinetic power in pc--scale radio jets, for a sample
of radio--loud quasars for which suitable data are available in the
literature. We find that the ionizing luminosity and the kinetic one are
of the same order of magnitude, suggesting that the processes responsible
for them are somehow related.  A strong magnetic field in equipartition
with the radiation field could be responsible for regulating both
processes.  BL Lac objects seem to follow a similar behaviour, but with
comparatively fainter broad line emission. 

\end{abstract} 


\section{Introduction}

A key ingredient in our understanding of the physics and energetics of
Active Galactic Nuclei (AGN), is the relation between the spectacular
phenomenon of jets and the accretion process and/or the power supply in
the central `engine'. The supposed presence of material in the form of an
accretion disc in the sources associated with jets, in forming stars,
galactic X--ray binaries and in powerful AGN, strongly suggests a
fundamental link between these two phenomena (e.g. Pringle 1993,
Lynden--Bell 1996 and references therein).  Direct observational evidence
for this connection has been revealed by the recent HST images of the jet
and disc of M87 (Ford et al. 1994). 

A possible approach to get insight into this link is to establish
correlations and dependences among observable/measurable quantities which
are likely to be connected to the accretion and jet processes. This
approach has been adopted by several authors, who have discussed the
relationship between luminosity in line emission, on different scales, and
radio and/or kinetic luminosity on large (kpc-Mpc) (e.g. Stockton \&
MacKenty 1987; Baum \& Heckman 1989a,b; Rawlings \& Saunders 1991 [RS91];
Miller, Rawlings \& Saunders 1993) and small (radio core) scales (Falcke,
Malkan \& Biermann 1995). Recently, using radio data on VLBI ($\sim$
parsec) scales and the standard synchrotron self--Compton theory, Celotti
\& Fabian (1993) [CF93] estimated the jet kinetic power to put constraints
on the matter content of jets.  Despite the fact that this method does not
directly face the fundamental questions of the mechanism(s) responsible
for the collimation and acceleration of jets, it still offers quantitative
constraints and reveals important clues. 

In the light of the above results, here we focus on the luminosity in
broad emission lines, and explore its relation with the kinetic one for a
sample of radio--loud objects for which the pc--scale kinetic power can be
estimated from VLBI data. The advantage of considering the luminosity in
broad lines, instead of narrow ones, as a (possibly isotropic) indicator
of the accretion power is threefold.  Firstly, there are indications that
the narrow line emission could be partly caused by photoionization induced
by shocks associated with jets or high velocity gas and therefore not
necessarily from the putative accreting material (e.g. Sutherland,
Bicknell \& Dopita 1993; Capetti et al. 1996).  Secondly, by looking for
information on broad line emission we are able to find more data compared
to the narrow line luminosities, and therefore increase the size of the
sample for which we can estimate the line luminosity. Finally, the
expected typical size of the broad line region is in fact roughly
comparable with the parsec scale.  This implies that we are more likely to
look at phenomena over the same temporal scale in the active life of the
nucleus.  We stress that the `isotropy' of broad line emission we refer to
is relative to objects which, according to the current unification
schemes, are observed within the opening angle of the putative obscuring
torus (see e.g. Urry \& Padovani 1995 for a recent review). 

The outline of the paper is as follows. In Section 2, we describe the
sample of sources, while in Sections 3 and 4 the methods applied to
estimate the luminosity in broad lines and the jet kinetic power are
outlined. In Section 5 we examine the case for considering also BL Lacs
objects in our study. The results are presented in Section 6 and discussed
in Section 7. Section 8 summarizes our conclusions. 

Cosmological parameters H$_0$=50 km/s/Mpc and q$_0$=0 have been adopted
throughout the paper. 

\section{The sample} 

We consider the sample of radio--loud AGN assembled by Ghisellini et al.
(1993) [G93], which includes 105 sources with VLBI angular diameter data
available in the literature, and for which it is therefore possible to
apply the synchrotron self--Compton (SSC) model and estimate the kinetic
luminosity in the mas (milliarcsec) scale jet. This is the only criterion
applied to select the sources, and therefore the sample is not, in any
sense, complete.  As already mentioned, we exclude from it radio galaxies,
which we expect to be observed at large angles with respect to the jet
axis and whose broad line emission is therefore likely hidden by obscuring
material. 

Among these radio--loud sources, we then consider the three subsamples of
broad line objects, namely core dominated High and Low Polarization
Quasars (HPQ and LPQ respectively) and lobe dominated quasars (LDQ),
obtaining a sample of 64 sources.  For a more detailed discussion about
their classification as well as further data and references we refer to
G93.  Fluxes at 1 keV for four sources without X-ray data in G93 are
derived from the {\it ROSAT} WGA catalogue (White, Giommi \& Angelini
1994) as described in Padovani, Giommi \& Fiore (1996). This translates
into a better determination of the Doppler factor (see below), which is
otherwise based, for sources without X-ray data, on the optical flux. 

\section{Broad line luminosity} 

We then have to estimate the total luminosity emitted in broad lines,
$\ll$. There is not any solidly established procedure to derive $\ll$ and
(unlike for example Falcke et al. 1995) we adopt the following method,
which involves three steps: i) define a set of lines which dominates the
total broad line emission;  ii) establish their relative flux ratios; iii)
extrapolate, using these ratios, from the luminosity in lines with
measured flux to the luminosity in all lines. 

i) Clearly, the more measured line fluxes are available, the more the
estimate is correct. Furthermore, the use of fluxes from several lines
increases the statistics in term of number of sources in the sample and
allows a better coverage in redshift. We consider fluxes for the following
lines: Ly$\alpha$, C~IV, Mg~II, H${\gamma}$, H${\beta}$, and H${\alpha}$,
which are amongst the major contributors to the total $\ll$ (making up in
fact about 60 per cent of it) and are well identified in quasar spectra. 
We then collect from the literature data on the broad line fluxes. In
order to obtain data as homogeneous as possible, and minimize the
uncertainties, we only consider values of line fluxes (or luminosities)
either when directly given or when the equivalent width and continuum flux
at the corresponding line frequency are reported by the same authors (for
small differences between the line and continuum frequencies, we
extrapolate the continuum, adopting a slope $\alpha_{\rm UV}=0.5$ between
2200 \AA\ and 2800 \AA\ , $F(\nu)\propto \nu^{-\alpha}$). 

\noindent We find line fluxes for a subsample of 43 of the 64 objects,
with an average of about two lines per source. When more than one value of
the same line flux was found in the literature we either considered the
most recent reference, which was likely to report values measured with
more accurate techniques, or estimate the arithmetic average of the flux
(in almost all the cases the difference was minimal). We have not applied
a reddening correction to the line fluxes, because the average data are
not corrected. 

ii) We use the line ratios reported by Francis et al. (1991) (see their
Table~1 for a complete list), mainly because of the large statistic and
the range of lines included \footnote {The lack of complete information on
the line spectra at all redshifts as well as some intrinsic scatter lead
to significant differences in the line ratios reported by different
authors. The ratios given (e.g. Boyle 1990, Netzer 1990, see also Baldwin
et al. 1995 and references therein) differ both for the number of lines
reported and for individual flux ratios.  In this sense, the adopted
procedure necessarily introduces an unavoidable dispersion. For example,
the line ratios recently reported by Zheng et al. (1996) from HST data,
lead to a difference of about 25 per cent in the total luminosity with
respect to Francis et al. (when the lines in common to the two works are
considered)}.  They refer to an optical sample, mostly consisting of radio
quiet sources, but no major differences in the broad line fluxes between
radio quiet and radio loud objects has been clearly determined (e.g.
Corbin 1992, see also Steidel \& Sargent 1991, Boroson \& Green 1992,
Wills et al. 1993; Zheng et al. 1996). 

iii) We then consider the sum of the line luminosities of all the lines
reported by Francis et al. (1991) with respect to the Ly$\alpha$, to which
we assign a reference value of 100 (hereafter the asterisk refers to
luminosities in the same units). To this, we also add the contribution
from $\lha^*$ (which is not included in the list of Francis et al.), with
a value of 77 (from Gaskell et al. 1981). This gives a total $\langle
\ll^* \rangle $ = 555.77$\sim$ 5.6 $\lla^*$. Therefore, given the sum of
the observed luminosities in a certain number of broad lines $\Sigma_{\rm
i} L_{\rm i, obs}$, the total $\ll$ can be calculated as

\begin{equation}
\ll = \Sigma_{\rm i} L_{\rm i,obs} \times {\langle \ll^* \rangle  \over 
                      \Sigma_{\rm i} L_{\rm i, est}^*} 
\end{equation}

\noindent where $\Sigma_{\rm i} L_{\rm i, est}^*$ is the sum of the
luminosities from the same lines, estimated through the adopted line
ratios. That is, the sum of the luminosities of the lines for which we
have data is scaled by the ratio of the total $\langle \ll^* \rangle$
divided by the sum of the luminosities of the same lines in the composite
spectrum of Francis et al. (1991). 

\noindent By using eq.~(1), we then derive $\ll$. Information on the lines
used, relative references and values of $\ll$ for each of the 43 objects
are reported in Tables~1a,b,c for the different classes of quasars. 

\section{Kinetic luminosity} 

The estimate of the kinetic luminosity is based on the adoption of the SSC
model, applied to the radio (VLBI) data and X--ray (or optical) fluxes.
The standard SSC theory allows one to derive limits on the comoving number
density of the emitting particles, $n$, the magnetic field intensity and
the relativistic Doppler factor, $\delta\equiv \Gamma^{-1} (1-\beta
\cos\theta)^{-1}$. Here $\Gamma$ is the Lorentz factor relative to the
bulk motion with velocity $\beta c$ of the emitting plasma, and $\theta$
is the angle between the direction of motion and the line of sight.  We
refer to CF93 and references therein for a complete description of the
(standard) method applied. 

Given $n$, $\delta$ and the VLBI angular size, the kinetic 
luminosity is then derived as

\begin{equation}
\lk\simeq \pi r^2_{\rm VLBI}\, n\, m_{\rm e} c^2\, \Gamma^2 \beta c
\qquad {\rm erg\, s}^{-1}
\end{equation}

\noindent for $\Gamma >> 1$, where $r_{\rm VLBI}=\theta_{\rm d} d_{\rm
L}/2/(1+z)^2$ is an estimate of the jet cross section from the measured
VLBI angular diameter $\theta_{\rm d}$ and the luminosity distance $d_{\rm
L}$ \footnote{We note that the estimate of the jet cross section could
take into account both a projection factor and the jet opening angle (e.g.
estimated from the relativistic Mach number). However, given the
uncertainties on the direction of the line of sight with respect to the
jet axis for the different classes of objects, and the fact that the
observed components do not show an elongated structure along the jet axis,
we assume that we are measuring the dimension of a quasi--spherical
component. Therefore we choose not to include these corrections (whose
product in any case results to be of the order of one)}, m$_{\rm e}$ is
the electron rest mass.  The adopted spectral index of the non--thermal
(synchrotron) radiation is $\alpha=0.75$.  The $\Gamma^2$ term in eq.~(2)
accounts for the particle energy flux ($\propto \Gamma\beta$) and the
transformation of the density ($n$ is the {\it comoving} number density). 

In the following, we present the main assumptions adopted in the estimate 
of $\lk$ through eq.~(2). 

As extensively discussed in CF93, the major uncertainties are: a) the
matter content of jets, i.e. whether it is made of an electron--proton
and/or an electron--positron pair plasma;  b) the extension of the
(differential) energy distribution of the emitting particles, $N(\gamma)
\simeq 2\alpha \gamma_{\rm min}^{-1} n (\gamma/\gamma_{\rm
min})^{-(2\alpha+1)}$, where $\gamma$ is the electron Lorentz factor, and
$\gamma_{min}$ therefore determines the total particle number density. 
Eq.~(2) apparently neglects both a contribution from heavy particles (i.e.
protons) and any possible advection term from either components.  However,
following the findings and discussion of CF93, the kinetic power seems to
be correctly estimated by that expression (see CF93 for further details).
Considering, for simplicity, only plasmas dominated either by an
electron--positron or electron--proton component, eq.~(2) corresponds to
the assumption of an electron--proton plasma with a low energy cutoff in
the (electron) distribution at around $\gamma_{\rm min} \sim 100$, or a
pair plasma in which the particles cool down to $\gamma_{\rm min}\sim 1$.
We do not include the possibility of a two-fluid jet (see e.g. Henri \&
Pelletier 1991). 

Another relevant point is the assumption about the derivation of the
Lorentz factor $\Gamma$ from the Doppler factor. As in CF93, we use the
minimum $\Gamma$ for any given $\delta$ [i.e. $\Gamma = 0.5(\delta +
1/\delta$)].  Lacking more accurate information, this seems to be the
wisest choice.  In fact, the other observational parameter, which enters
in the estimate of $\lk$, is the X--ray (or optical) flux, which we assume
is all due to the SSC process. An overestimate of this flux, $F_{\rm x}$,
would lead to an overestimate of the kinetic luminosity.  In fact the
estimate of the particle density $n\propto F_{\rm x}$, while the estimate
of the Doppler factor $\delta\propto 1/F_{\rm x}^{1/(4+2\alpha)}$,
yielding $\lk \propto F_{\rm x}^{(1+\alpha)/(2+\alpha)}$. Therefore the
choice of a minimum $\Gamma$ would tend to reduce the total uncertainty in
the estimate of $\lk$. It is worth noticing that the X--ray flux can be
particularly overestimated for sources where clearly the synchrotron
component still dominates at X--ray energies (fluxes used here are
estimated at 1 keV). 

The only difference in the computation of $\lk$ with respect to CF93 is
the treatment of sources with estimated $\delta<1$. In fact, because here
we are interested in an average value of $\lk$, rather than upper limits
on it, for sources with $\delta<1$ we derive $\Gamma$ from an average
$\delta$ (while CF93 assumed $\Gamma^2 \beta = 1$). In agreement with
current unification schemes (e.g. Urry \& Padovani 1995), we consider the
average $\delta$ of the core dominated quasars (or BL Lacs), as derived by
G93 ($\langle \delta \rangle \simeq$ 6 and 3, respectively), which should
better correspond to the intrinsic Lorentz factor.  As shown in Table~1,
the 13 sources with $\delta<1$ are mainly LDQ, and make up about 24 per
cent of the sample.  In any case, these objects are considered separately
in our analysis. 

\section{The case of BL Lac Objects} 

In order to further explore the connection between $\lk$ and $\ll$, and in
view of the recent findings of (weak) broad emission lines in BL Lac
objects (Stickel, Fried \& K\"uhr 1993b; Vermeulen et al. 1995), we also
estimated the above quantities for 12 BL Lacs belonging to the G93 sample.
However these sources have been treated somehow separately in all the
analysis. In fact, in agreement with current unification schemes, while
the three classes of quasars should be intrinsically identical, BL Lacs
are likely to represent a separate class. 
 
The list of BL Lac objects as well as the estimated luminosities are
reported in Table~1d \footnote{Note that to estimate the BLR flux of BL
Lac itself we have not used the recent observations of strong, broad
Balmer lines of Vermeulen et al. (1995), as most likely they represent an
atypical state of the object (see discussion in Vermeulen et al. 1995).}.
As discussed below, an important point to be stressed is that for this
class of sources the estimated $\ll$ is an upper limit. 

\section{Results}

\subsection{Luminosity correlations} 

Fig.~1 shows $\lk$ vs. $\ll$ for the different classes.  As one can see
from the spread in the data, when all sources of the current sample are
considered, there is only a weak statistical indication of a linear
correlation between the two luminosities, at a significance level of 95.8
per cent. An even weaker hint of correlation is present when individual
classes of sources or only sources with $\delta>1$ are considered. An
exception is constituted by the class of BL Lacs (with $\delta>1$), whose
kinetic and BLR luminosities appear to be correlated at the 98.4 per cent
level. 

An interesting point is that the luminosities show the same general trend
of the distribution reported by CF93 for the Narrow Line emission. 
However a direct comparison of these two sets of data is rather difficult
because it would require the inclusion of different covering factors in
order to ``rescale" the line luminosities. 

As can be seen in Fig.~1, the spread in the correlation is quite large,
plausibly due to intrinsic dispersion, but also to the significant
uncertainties of our assumptions. In order to ascertain the likely causes
of this spread we have first looked for a possible relation between the
core radio luminosity ($\nu_{\rm r} L_{\rm r,core}$) and $\ll$ and indeed
found that the two quantities appear to be highly correlated at the $>
99.9$ per cent level (note however that part of this strong correlation is
due to the common redshift dependence).  As shown in Fig.~2, the two
luminosities appear to be of the same order of magnitude, with the
expected tendency for the LDQ to have relatively lower core radio power. 

We therefore examined the possible spread introduced by the other
quantities on which the estimate of $\lk$ is based, namely $\theta$ and
$F_{\rm x}$. In order to do this, we assumed that in turn these two
quantities have no intrinsic spread: we found that, while the assumption
of a constant $\theta$ does not lead to a better correlation, the
tentative hypothesis of an intrinsic relation of $F_{\rm x}$ with the core
radio flux $F_{\rm r,core}$, $F_{\rm x}\propto F_{\rm r,core}$, gives
values of $\lk$ highly correlated with $\ll$ at the $99.9$ per cent level. 
This supports the view that the large scatter plausibly reflects a scatter
in the observed $F_{\rm x}$. Indeed, we expect that the observed strong
variability in the X--ray band, as well as the possible contribution of
emission not due to SSC (see e.g. Sikora et al.  1994), would introduce
large uncertainties. \footnote{We note however that, because of the weak
dependence of $\delta$ on $F_{\rm x}$, the estimates of the Doppler
factors would be only slightly affected (see also Section 4)}

Because of the large scatter found and the corresponding weak correlation,
we instead consider the ratio between the kinetic and broad line
luminosities to determine if this shows any interesting average property.
In Fig.~3, the distribution of this ratio is shown for the different types
of objects. Shaded areas refer to sources with an estimated $\delta<1$. As
already discussed, the distributions are quite broad, but the average (and
median) of this ratio for quasars is $\approx$ 10. The distributions for
HPQ, LPQ, and LDQ are consistently similar according to a
Kolmogorov-Smirnov (KS) test.  Adopting a typical BLR covering factor of
$\sim$ 0.1 (e.g. Netzer 1990), this implies that $\lk \sim \li$, the
ionizing radiation. 

This result is certainly intriguing, given: 
a) the uncertainties in the estimates of $\ll$ and especially $\lk$; 
b) the fact that they are based on completely independent calculations 
(where the only parameter in common is the red shift); 
c) the sample used, which does not have any characteristic
of completeness or clear selection criteria. 

In Table~2, we report the results on the average values (and medians) of
$\log(\lk/\ll)$ with their dispersions, for the different types of
sources. It can be seen that the same result on the ratio $\lk/\li$
roughly holds separately for the three classes of quasars (the ratio is
slightly higher for LDQ when all sources are considered but goes down
considerably for the objects with $\delta > 1$). 

It is worth noticing that the choice of estimating $\ll$ instead of trying
to directly measure $\li$ from the optical/UV/X--ray fluxes, is mainly due
to the difficulty in disentangling the non--thermal anisotropic radiation
from the more isotropic one (the `blue bump') in the observed flux (as
well as to the paucity of IUE data for these sources). We would in fact
expect the ratio of these two components to be strongly dependent on the
observation angle, as discussed below (see also Browne \& Murphy 1987;
Wills et al. 1993). 

\subsection{BL Lac objects versus quasars}

Figures 1 and 3 and Table~2 show that $\lk/\ll$ for BL Lacs is much larger
than for quasars. Moreover, a KS test shows that the $\lk/\ll$
distribution for BL Lacs is different from that of quasars at the 99.1 per
cent level, independently of the exclusion of sources with $\delta < 1$
(but is consistent with that of LDQ only, which however include only very
few objects, especially if $\delta > 1$ is required). This difference is
reinforced by the fact that for BL Lacs this ratio should be considered as
a lower limit.  In fact, due to the lack of better information, $\ll$ has
been derived using the same line ratios adopted for quasars, as if all
those lines were also present (but very broad) in the spectrum of BL Lacs.
Moreover, in many sources absolutely no emission line is detected:
inclusion of these objects would have increased further the mean ratio
$\lk/\ll$. 

\section{Discussion} 

The main finding of this work is the approximate `equivalence' between
$\lk$ and $\li$. 

Intriguingly, a comparison of our results with those found in previous
works (e.g. RS91) about the correlation between the Narrow Line luminosity
and the power required to feed the extended lobes, shows that the same
behaviour seems to hold in an analogous way (when a different covering
factor is assumed) for the luminosity both in broad and in narrow lines.
Also in the latter case, in fact, one obtains an approximate equivalence
between the ionizing radiation and the kinetic power supplied to the lobes
or to the extended radio structures. 

The first implication of these findings is to reinforce and stress the
fact that the channeling of material into jets is indeed a powerful
process in term of the total energy budget of radio--loud AGN, and more
precisely this is comparable with the amount of the radiative power
emitted isotropically. In this sense, it is relevant to point out that, in
the most accepted scenario on the physical processes operating in the
central regions of AGN, $\ll$ is related to a (semi)isotropic and unbeamed
radiation.  Both luminosities would be therefore unaffected by beaming and
represent `intrinsic' powers.  As shown in Fig.~1, the energy output can
reach up to 10$^{48}$ erg s$^{-1}$.

Furthermore, if $\li$ is indeed related to the power dissipated during the
accretion phase, this finding would suggest that there is a close link
between the accretion process and the phenomenon of jets, or,
alternatively, that a common `element' regulates the amount of luminosity
radiatively dissipated during the accretion process and the power
channeled into the jet in kinetic form.  This `element' should be indeed
tightly linked to the above quantities if it can produce a significant
correlation between the observed variables. 

This deep link has also been suggested by Saunders (1991) on the basis of
the observational similarity found between the luminosity in narrow line
and the kinetic power, estimated on the large radio scales (RS91; see also
Falcke et al. 1995, Wills \& Brotherton 1995).  An interesting possibility
is that the key role is played by magnetic fields. 

\subsection{The possible role of magnetic fields}

On one side, recent theories of disc accretion and observations at high
energies favor the idea that most of the accreted power is dissipated in a
corona above the disc, due e.g. to magnetic reconnection.  On the other
hand, the most accredited possibility for the collimation and acceleration
of jets involves the presence of magnetic fields. In fact this hypothesis
accounts for different aspects of the physics of jets, as well as
different environments for their formation.  The power would be
`extracted' from the rotational/gravitational energy of an accretion disc
or the black hole itself, and then channeled as kinetic power and/or
Poynting flux (see e.g. the reviews by Blandford 1993, Spruit 1996 and
references therein). 

If the magnetic field indeed plays an important part in both processes,
then one could expect some sort of correlation between the energy
dissipated and the kinetic power.  In other words, the correlation found
here could be the signature of the equivalence of the power generated
through accretion and the one extracted from the disk/black hole
rotational energy and converted into kinetic form. The direct comparison
of the predictions of this models with observational data is still quite
limited, and substantially concerns jets associated with star-size
objects. In the case of AGN, we can only perform a rough estimate,
assuming that the magnetic field is in equipartition with the radiation
field, which plausibly dominates the pressure in the inner region of an
accretion disk, and that a similar efficiency regulates the conversion of
the produced energy into radiation and bulk energy. This estimate is
consistent with obtaining similar values for the kinetic and ionizing
power.  For example, for an isolated Kerr black hole, the spin energy
which can be extracted in electromagnetic form is approximately given by
(e.g. Blandford 1990)

\begin{equation}
\lr \simeq 10^{45} \left( {a\over m} \right)^2 B^2_4M_8^2
\qquad {\rm erg\, s}^{-1}
\end{equation}

\noindent where $ac$ is the specific angular momentum and $m=GM/c^2$ the
gravitational radius ($a < m$).  Then for a rapidly rotating black hole

\begin{equation}
\lr \sim 8\pi\times 10^{37} U_{\rm B} M_8^2 \sim 2\times 10^{37} 
\left({L_{\rm acc}\over c R_{\rm s}^2}\right) M_8^2 \sim 0.8 L_{\rm acc}
\end{equation}

\noindent In this estimate equipartition between the magnetic ($U_B$) and
the radiation energy density ($U_{\rm rad}\simeq L_{\rm acc}/4\pi c R_{\rm
s}^2$) has been assumed, with a typical size for the region of the order
of a Schwarzschild radius, $R_s$.  If these luminosities are converted
with similar efficiency into kinetic and ionizing luminosities,
respectively, then $\lk \sim \li$. Clearly, the above evaluation is based
on gross approximations and physical hypothesis, which currently are still
at the level of speculation (e.g. Spruit 1996 and references therein). 

It is worth noticing that if energy is extracted from a spinning objects,
it can propagate in electromagnetic and/or kinetic form. Here we are
indeed assuming that this power ultimately, and on scales smaller than a
parsec, is largely converted into kinetic power (from a theoretical point
of view, this balance depends on the boundary conditions of the
hydromagnetic problem, see e.g. Begelman 1994 and references therein). It
is then conserved as such up to much larger scales, as shown by the
comparison with the results of RS91 and the fact that radiative
dissipation within the jet itself (unbeamed radiation) is a negligible
fraction of the estimated $\lk$ (CF93). Further support to this
possibility comes from the recent results by Bowman, Leahy \& Komissarov
(1996), who show that even dissipation caused by entrainment of material
in the jet and the consequent deceleration, cause only a relatively small
loss in kinetic power. 

Alternatively, one could argue that the found similarity between $\lk$ and
$\li$ supports the hypothesis that the acceleration of the jet is indeed
caused by the radiation field generated in the disc.  Our main objection
to this point is the inefficiency of radiative acceleration to produce
highly relativistic flows, even in the most favourable situation where the
momentum is transferred through synchrotron absorption (G93). Our findings
would indeed require a very high, close to $\sim $ 100 per cent,
efficiency in the conversion of radiative to kinetic luminosity. 

\subsection{Anisotropic ionizing continuum?}

In what discussed it has been assumed that the radiation ionizing the
broad line emitting gas is generated (semi)isotropically by the plasma
accreting onto the central object.  In fact, a further possible
interpretation of the results is that another anisotropic ionizing source,
namely the beamed radiation from the jet itself, could be responsible for
ionizing a significant fraction of the BLR emitting material (see e.g. 
Wilson 1993 on photoionization of the narrow line emitting gas). While
this has been suggested as a possible explanation of the relatively rapid
line variability observed in some radio--loud objects (e.g. P\'erez,
Penston \& Moles 1989; see also Ghisellini \& Madau 1996), it would raise
the problem of understanding the similarity of the broad line component in
radio--quiet and radio--loud sources, the former ones lacking evidence of
a powerful (and beamed) jet. Furthermore, if the beamed continuum
dominates the ionizing flux, then misaligned objects should systematically
have broad lines of much larger equivalent widths than observed. 

In this respect, it is also interesting to note that, while the luminosity
itself depends on the class of sources considered (namely HPQ, LPQ and
LDQ), in agreement with the most popular unification models, still the
typical ratios between kinetic and broad--line luminosities appear to be
independent of the type of object. This does indeed support the fact that
$\ll$ is most likely to be an orientation independent parameter (for
sources viewed down the ``obscuring torus'').  One could even think of
determining, by measuring $\li$ directly through optical/UV/X--ray data,
how the ratio of isotropic (as derived from $\ll$) versus anisotropic
radiation varies with different objects, as a further orientation
parameter. As an example, an estimate of this kind for $\li/\ll$ (by using
the values of $\li$ derived by Padovani \& Rafanelli 1988 and Padovani
1989) gives a ratio of about 10 for 3C 120, 3C 273, 3C 390.3, and PKS
1510$-$089 but about 100 for a highly beamed source like 3C 279. Finally
this ratio could be also expected to have some correlation with the degree
of linear polarization (e.g. Wills 1991). 

\subsection{BL Lac objects}

One could interpret the results for the class of BL Lac objects as due to
a paucity of $\ll$, compared to quasars. This can derive from inefficient
dissipation during accretion and/or a smaller accretion rate, and
therefore a lower $\li$. This possibility is also reinforced by the fact
that our estimate of the broad line emission in BL Lacs, as previously
discussed, tends to overestimate $\li$. If Fanaroff-Riley type I radio
galaxies (FR I: Fanaroff \& Riley 1974) are the parent population of BL
Lacs, as currently believed (e.g. Urry \& Padovani 1995), this would also
be an independent piece of evidence in agreement with the findings of
Baum, Zirbel \& O'Dea (1995). These authors, by studying the optical and
radio properties of FR I and FR II radio galaxies (the latter thought to
be the parent population of radio quasars), have in fact shown that, at
the same host galaxy magnitude or radio luminosity, FR Is have an order of
magnitude weaker line emission than FR IIs. However, it is also possible
that the lower $\ll$ is caused by a lack of reprocessing broad line
material. 

Alternatively, one could argue that the $\lk/\li$ ratio is similar for BL
Lacs and quasars, but that eq.~(2) overestimates $\lk$ in the former
objects.  In particular, an intriguing possibility is that jets in BL Lacs
and their 'parent population' (namely FR I radio galaxies) are mainly
composed of an electron--positron plasma, while jets in quasars have a
dominant electron--proton component, even if the number density of
emitting particles is the same in the two classes of objects. This would
reduce $\lk$ in BL Lacs with respect to quasars, and may indeed be
responsible for the different kinetic power and possibly the different
morphology of FR I radio galaxies compared to their powerful counterparts.
\footnote{Given the continuity of radio properties of these two classes of
sources, we would expect a rather continuous change in the composition of
the jet.} Such a hypothesis has been indeed formulated very recently by
Reynolds et al. (1996), on the basis of a detailed study of the dynamics
and radiative properties of the jet in M87, and will be further
investigated elsewhere (Bodo et al., in preparation). 

\section{Conclusions} 

We have estimated the kinetic luminosity in the pc--scale radio jets of a
sample of radio--loud AGN, by applying the standard SSC theory and
compared it to the luminosity observed in broad lines. This has shown a
quantitative similarity in quasars between these two forms of power, which
could be translated into the equivalence between the ionizing luminosity,
which is most likely the luminosity radiatively dissipated during the
accretion process, and the luminosity channeled into jets in kinetic form.
BL Lacs, however, show a deficit of $\ll$ or, alternatively, an excess of
$\lk$. 

We find only a weak hint of correlation between $\lk$ and $\ll$ which
would extend over $\sim$ 5 orders of magnitude and, when a different
covering factor is taken into account, confirms the analogous behaviour of
the Narrow Line luminosity with the power needed to feed the extended
radio lobes (e.g. RS91), and the kinetic jet power (CF93).  This is likely
to be of the same nature of similar correlations between the radio
luminosity and the emission of narrow lines on extended scales (e.g. Baum
\& Heckman 1989a,b). The large scatter we find in the correlation is
probably due to the non-simultaneous radio and X-ray data which enter in
our calculation, as $\lk$ depends rather strongly on $F_{\rm x}$. 

In any case, the similarity of $\lk$ and $\li$ seems to indicate that a
common factor regulates both the luminosity dissipated during the
accretion phase and the kinetic power of the jet. This clearly does not
univocally define the factor responsible, however we suggest that this can
be a piece of evidence in favour of the key role of magnetic fields. 

It is also interesting to point out that the estimate of the broad line
luminosity can have important predictive power on the models for the
generation of the X-- and especially $\gamma$--ray radiation in different
classes of blazars (see e.g. the review by Sikora 1994) which stress the
importance of Comptonization of the diffuse radiation field through which
the jet propagates.  It should be however noticed that here the estimates
of the jet Doppler factor implicitly assume that the X--ray emission is
mostly produced by the SSC mechanism.  Polarization information in the
X--ray band (e.g. Celotti \& Matt 1994) will be therefore extremely
important to constrain the emission models. 

\section*{Acknowledgments} 

We thank the referee, Heino Falcke, for helpful and constructive comments
which helped improving the paper.  AC acknowledges the MURST for financial
support. This research has made use of the NASA/IPAC Extragalactic
Database (NED), which is operated by the Jet Propulsion Laboratory,
California Institute of Technology, under contract with the National
Aeronautic and Space Administration. 

\section*{References} 

\refitem Baldwin J.A., Wampler E.J., Burbidge E.M., 1981, ApJ, 243, 76

\refitem Baldwin J.A., Wampler E.J., Gaskell C.M., 1989, ApJ, 338, 630

\refitem Baldwin J.A., Ferland G., Korista K., Verner D., 1995, ApJ, 455,
L119

\refitem Baum S.A., Heckman T.M., 1989a, ApJ, 336, 681

\refitem Baum S.A., Heckman T.M., 1989b, ApJ, 336, 702

\refitem Baum S.A., Zirbel E.L., O'Dea C.P., 1995, ApJ, 451, 88 

\refitem Begelman, M.C., 1994, in The Nature of Compact Objects in Active 
Galactic Nuclei, Robinson, A., Terlevich, R.J., eds., Cambridge 
University Press, p.~361

\refitem Bergeron J., Kunth D., 1984, MNRAS, 207, 263 

\refitem Blandford R., 1990, in Active Galactic Nuclei, 20$^{th}$
SAAS--FEE Course, Courvoisier, T.J.-L., Mayor, M., eds. (Springer--Verlag)

\refitem Blandford R., 1993, in Astrophysical Jets, Burgarella D., Livio
M., O'Dea C., eds. (Cambridge University Press), p.~15

\refitem Boroson, T.A., Green, R.F., 1992, ApJS, 80, 109

\refitem Bowman, M., Leahy, J.P., Komissarov, S.S., 1996, MNRAS, 279, 899

\refitem Boyle B.J., 1990, MNRAS, 243, 231

\refitem Browne I.W.A., Murphy D.W., 1987, MNRAS, 226, 601

\refitem Capetti A., Axon D.J., Macchetto F., Sparks W.B., 
Boksenberg A., 1996, ApJ, 454, 85

\refitem Celotti A., Fabian A.C., 1993, MNRAS, 264, 228 (CF93)

\refitem Celotti A., Matt G., 1994, MNRAS, 268, 451

\refitem Clavel J., Wamsteker W., 1987, ApJ, 320, L9

\refitem Corbin M.R., 1992, ApJ, 391, 577

\refitem Falcke, H., Malkan, M.A., Biermann, P.L., 1995, A\&A, 298, 375

\refitem Fanaroff B.L., Riley J.M., 1974, MNRAS, 167, 31p 

\refitem Ford H.C., et al., 1994, ApJ, 435, L27 

\refitem Francis P.J., Hewett P.C., Foltz C.B., Chaffee F.H., 
Weymann R.J., Morris S.L., 1991, ApJ, 373, 465

\refitem Gaskell C.M., Shields G.A., Wampler E.J., 1981, ApJ, 249, 443 

\refitem Ghisellini G., Madau P., 1996, MNRAS, 280, 67

\refitem Ghisellini G., Padovani P., Celotti A., Maraschi L., 1993, 
ApJ, 407, 65 (G93)

\refitem Gondhalekar P.M., 1990, MNRAS, 243, 443

\refitem Henri, G., Pelletier, G., 1991, ApJ, 383, L7

\refitem Jackson N., Browne I.W.A., 1991, MNRAS, 250, 414

\refitem Junkkarinen V.T., 1984, PASP, 96, 539 

\refitem Lynden--Bell D., 1996, MNRAS, 279, 389

\refitem Marziani P., Sulentic J.W., Calvani M., Perez E., Moles M.,
Penston M.V., 1993, ApJ, 410, 56

\refitem Marziani P. Sulentic J.W., Dultzin-Hacyan D., Calvani M., Moles M., 
1996, preprint

\refitem Miller, P., Rawlings, S.G., Saunders, R.D.E., 1993, MNRAS,263, 425

\refitem Morganti R., Ulrich M.-H., Tadhunter C. N., 1992, MNRAS, 254, 546 

\refitem Netzer H., 1990, in Active Galactic Nuclei, 20$^{th}$ SAAS--FEE
Course, Courvoisier, T.J.-L., Mayor, M., eds. (Springer--Verlag)

\refitem Netzer H., et al., 1994, ApJ, 430, 191 

\refitem Neugebauer G., Oke J.B., Becklin E.E., Matthews K., 1979, ApJ, 230, 79

\refitem Oke J.B., Shields G.A., Korycansky D.G., 1984, ApJ, 277, 64

\refitem Osmer P.S., Porter A.C., Green R.F., 1994, ApJ, 436, 678 

\refitem Padovani P., 1989, A\&A, 209, 27

\refitem Padovani P., Rafanelli P., 1988, A\&A, 205, 53

\refitem Padovani P., Giommi P., Fiore F., 1996, MNRAS, in press

\refitem P\'erez E., Penston M. V., Moles M., 1989, MNRAS, 239, 75

\refitem Pringle J., 1993, in Astrophysical Jets, Burgarella D., Livio M.,
O'Dea C., eds. (Cambridge University Press), p.~1

\refitem Rawlings S.G., Saunders R.D.E., 1991, Nature, 349, 138 (RS91)

\refitem Reynolds C.S., Fabian A.C., Celotti A., Rees M.J., 1996, 
MNRAS, in press

\refitem Richstone D.O., Schmidt M. 1980, ApJ, 235, 361

\refitem Rudy R.J., 1984, ApJ, 284, 33

\refitem Saunders R.D.E., 1991, in Extragalactic Radio Sources -- From
Beams to Jets, 7$^{th}$ IAP Meeting, Roland J., Sol H., Pelletier G., eds. 
(Cambridge University Press), p.~344

\refitem Scarpa R., Falomo R., Pian E., 1995, A\&A, 303, 730

\refitem Sikora M., 1994, ApJS, 90, 923

\refitem Sikora M., Begelman, M., Rees, M.J., 1994, ApJ, 421, 153

\refitem Sitko M.L., Junkkarinen V.T., 1985, PASP, 97, 1158 

\refitem Smith H.E., Spinrad H., 1980, ApJ, 236, 419

\refitem Spruit H.C., 1996, in Physical processes in Binary Stars, Wijers
R.A.M.J., Davies M.B., Tout C.A., eds., NATO ASI Series (Kluwer
Dordrecht), in press

\refitem Steidel C.C., Sargent W.L.W., 1991, ApJ, 382, 433

\refitem Steiner J.E., 1981, ApJ, 250, 469

\refitem Stickel M., K\"uhr H., 1993a, A\&AS, 100, 395 

\refitem Stickel M., K\"uhr H., 1993b, A\&AS, 101, 521 

\refitem Stickel M., Fried J.W., K\"uhr H., 1989, A\&AS, 80, 103 

\refitem Stickel M., Fried J.W., K\"uhr H., 1993a, A\&AS, 97, 483

\refitem Stickel M., Fried J.W., K\"uhr H., 1993b, A\&AS, 98, 393

\refitem Stockton A., MacKenty J.W., 1987, ApJ, 316, 584

\refitem Sutherland R.S., Bicknell G.V., Dopita M.A., 1993, ApJ, 414, 506

\refitem Urry M.C., Padovani P., 1995, PASP, 107, 803

\refitem Vermeulen R.C., Ogle P.M., Tran H.D., Browne I.W.A., Cohen M.H., 
Readhead A.C.S., Taylor G.B., Goodrich R.W., 1995, ApJ, 452, L5 

\refitem White N.E., Giommi P., Angelini L., 1994, IAU Circ. 6100

\refitem Wills B.J., 1991, in Variability of AGN, Miller H.R., Wiita 
P.J., eds. (Cambridge University Press), p.~87

\refitem Wills B.J., Brotherton M.S., 1995, ApJ, 448, L81

\refitem Wills B.J., et al., 1983, ApJ, 274, 62

\refitem Wills B.J., Brotherton M.S., Fang D., Steidel C.C., Sargent W.L.W., 
1993, ApJ, 415, 563

\refitem Wills B.J., et al, 1995, ApJ, 447, 139

\refitem Wilson A.S., 1993, in Astrophysical Jets, Burgarella D., Livio
M., O'Dea C., eds. (Cambridge University Press), p.~121

\refitem Yee H.K.C., 1980, ApJ, 241, 894 

\refitem Zheng W., Kriss G.A., Telfer R.C., Grimes J.P., Davidsen A.F., 
1996, ApJ, in press

\vfill\eject

\section*{Figure captions} 

\noindent {\bf Fig.~1} The kinetic luminosity of pc--scale jets, as
estimated from the SSC theory, is plotted against the estimated luminosity
in broad lines, for different classes of radio--loud sources, namely Core
Dominated High Polarization Quasars ({\it full circles}), Core Dominated
Low Polarization Quasars ({\it open circles}), Lobe Dominated Quasars
({\it open squares}), and BL Lac objects ({\it crosses}). The underlined
symbols refer to sources for which the derived Doppler factor is less than
one (see text). The dashed line corresponds to $\lk = \ll$. 

\vskip 0.5 truecm \noindent {\bf Fig.~2} The core radio luminosity,
$\nu_{\rm r} L_{\rm r,core}$ vs the luminosity in broad lines, for the
same objects of Fig.~1.  The dashed line corresponds to $\nu_{\rm r}
L_{\rm r,core} = \ll$. While the correlation between the two luminosities
is largely due to the common redshift dependence, it still suggests that
the large spread in the correlation $\lk$ vs $\ll$ is due to uncertainties
in the estimate of $\lk$. 

\vskip 0.5 truecm \noindent {\bf Fig.~3} The distribution of $\log
(\lk/\ll)$ for the various classes of sources, that is (from top to
bottom): Core Dominated High Polarization Quasars, Core Dominated Low
Polarization Quasars, Lobe Dominated Quasars, all Quasars, and BL Lac
objects. Dashed areas indicate objects with an estimated $\delta <1$ (see
text). 

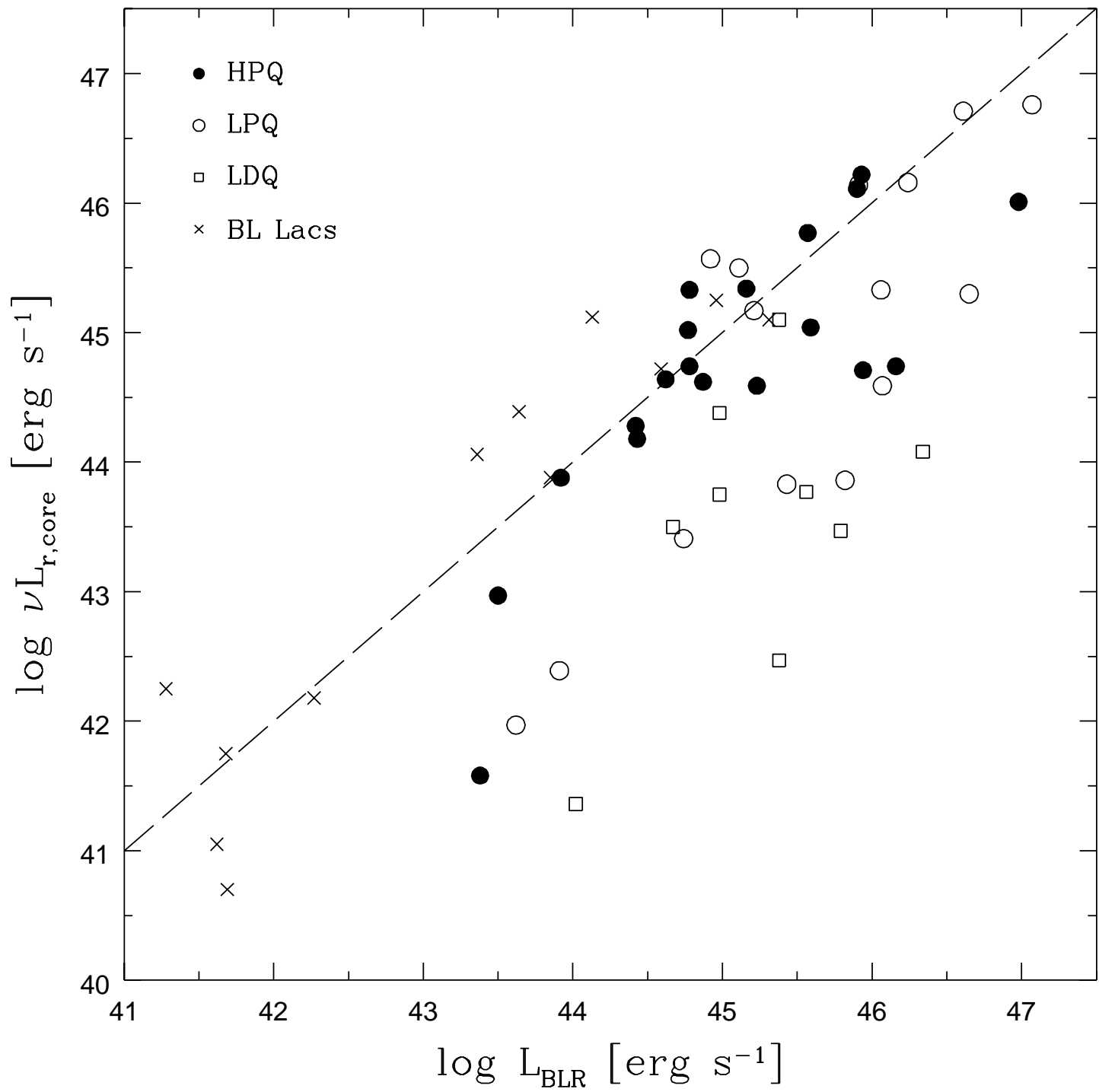
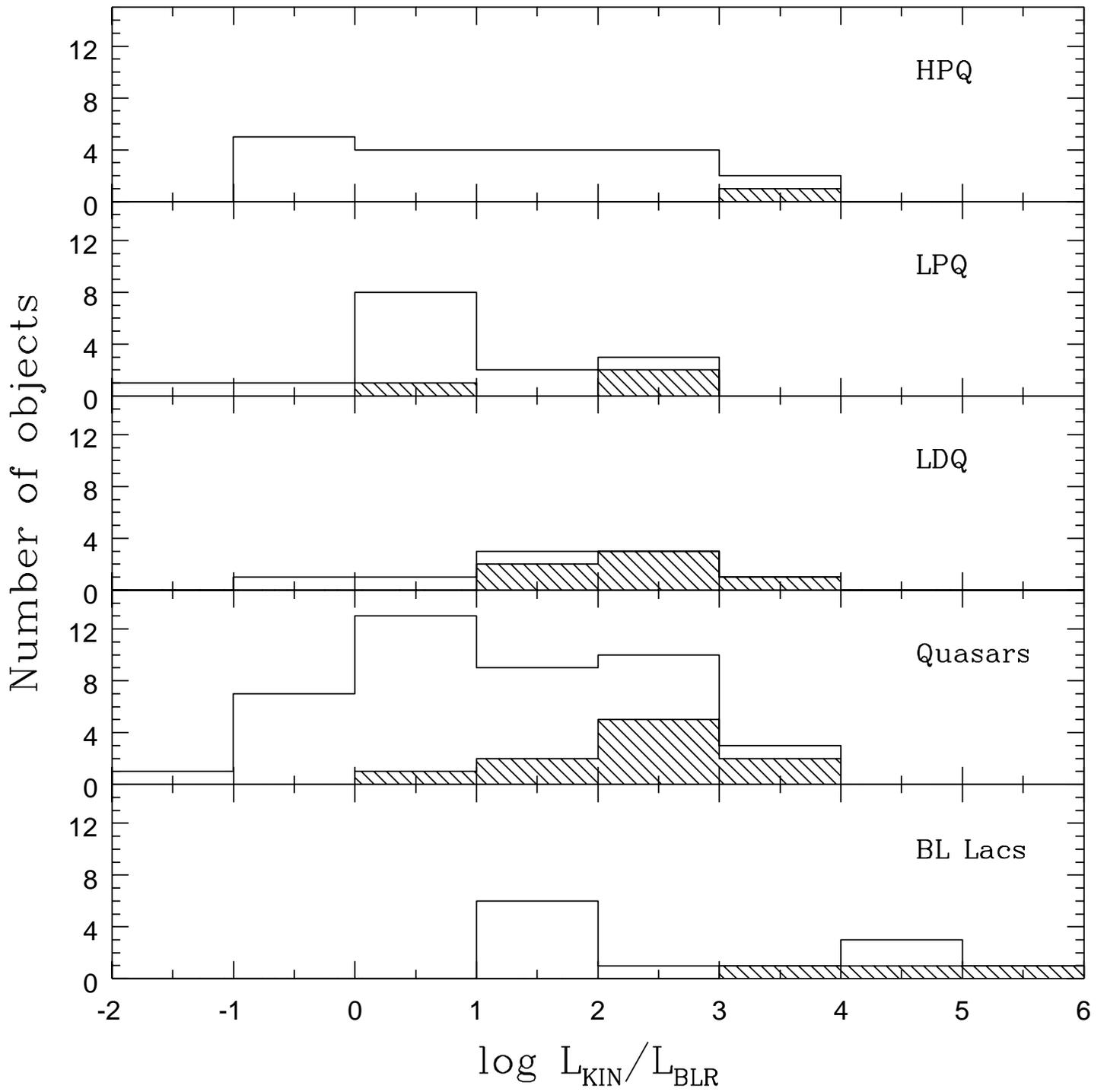
\end{document}